\documentclass{article}

\begin{document}

{\bf HOLOGRAPHY IN THE EPRL MODEL.}

\bigskip

by: Louis Crane, Mathematics department, KSU.

\bigskip

{\bf ABSTRACT:} {\it In this research announcement, we propose a new 
interpretation of the EPR quantization 
of the BC model using a functor we call the time functor, which is the first 
example of a CLa-ren functor. Under the hypothesis that the universe is in 
the Kodama state, we construct a holographic version of the model. 
Generalisations to other CLa-ren functors and connections to model category
theory are considered.}

\bigskip

{\bf Introduction. Mathematics and Physics of the new model }

\bigskip

In \cite{EPR},  \cite{LS} \cite{EP} and \cite{B}, a new version of the
BC model has been developed. The problems relating to the Hilbert
space and the geometrical interpretation have been resolved.

The purpose of this paper is to describe the mathematical form of the
new model, connect it to algebraic constructions of 3D TQFTs, and
thereby to make a new interpretation of the model.

There are several things we would want from a quantum theory of
gravity. It has not yet been demonstrated how to get them from the new
model. We list some critical physical problems, and explain what is
still missing to understand them.

\bigskip

1. We would like to see classical general relativity emerge for
distance scales above the Planck length. It was shown in \cite{B} that
classical geometry dominates the integral expression corresponding to
a classical four simplex in the limit of large spins. However there is nothing
in the model corresponding to larger composite ``coarse grained''
simplices, and no reason to associate higher spins to them.

\bigskip

2. A finite unambiguously defined theory should be possible. While the
new model is finite on individual simplicial complexes, there is no
procedure for choosing a particular complex to correspond to a
physical region. Summing over complexes ala GFT returns us to a formal
divergent expression, which is tragic after finiteness has come to us
in such a magical way.

\bigskip

3. It should be possible to define the theory on a bounded region of
spacetime. The Bekenstein bound should emerge. Adding boundaries to
the spatial region has not been considered. Quantum theory has its
foundation in an analysis of real experiments, which take place in
finite laboratories and do not take forever.

\bigskip

The specific approach we will take is to realize the Bekenstein bound
for the EPRL model on a bounded region by applying techniques from TQFT.

From a different perspective, we could hope to see the mathematical problem of the
nature of the continuum resolved by a quantum theory of gravity. Our
ideas of the continuum are the result of our experience with the
geometry of nature. The assumption that a continuum is infinitely
divisible has led Mathematics to a crisis of its foundation.

It is therefore quite plausible that an adequate quantum mechanical
study of the description of spacetime could replace the naive idealized
extension of classical mechanics to arbitrarily small distances on
which our current understanding of the real number system is based.

Later in this paper we shall argue that the physical problems
listed above and the foundational mathematical problem of the
continuum are naturally related, and furthermore, that the
mathematical form of the EPRL model provides an avenue for solving them all
at once.

In particular we will show that the classical limit problem is closely related
to the mathematical field of model categories, which is a well studied
and highly successful approach to replacing point sets as the
foundation of topology.

In order to investigate this we begin by describing the structure of
the new model in its general mathematical form, which has not been
done by the physicists who developed it. Generalisations of
the new model may also prove important in extending the theory to
include matter fields.

\bigskip

{\bf The Time Functor}

\bigskip

In what follows we will consider only the physical Lorentz
signature version of the model.

The point of departure for the new model is the discovery of an
elegant way to impose the simplicity constraints of the BC model
weakly rather than strongly. This amounts to reducing the irreducible
representations of the Lorentz group which are the building blocks of
the theory into sums of representations of a suitable copy of SU(2)
and selecting only the lowest spin representation which appears. 

Since
an important aspect of this procedure is that it relates the
Hamiltonian picture of loop quantum gravity, in which states are
described by spin networks in space, with relativistic spin networks
in spacetime; we want to think of it as a rule which assigns `` spacetime''
representations of SL(2,C) to ``space'' representations of
SO(3,R). The proper mathematical expression of this is a functor 
\bigskip

$ F_{\gamma } : REP(SO(3,R)) \rightarrow REP (SL(2,C)); $ 

\bigskip

 which depends on the Immirzi parameter $\gamma $. The case $\gamma =0 $
is a degenerate case in which much of what follows is incorrect. 

This functor assigns to each irreducible representation $R_k$ of
SO(3,R) the irreducible $R(k, \gamma k)$ of SL(2,C). Since the only
morphisms between irreducibles in either category are multiples of the
identity, the action of the functor on morphisms is immediate.

Considerably more subtle are the tensor and ``renormalizability''
properties of this functor, which are generalizations of the very
special facts which make the Lorentzian EPRL model finite and
physically interesting.

The action of the time functor on direct sums is straightforward. The
behavior on tensor products is more subtle. The image under $F_\gamma
$ of the tensor product of two objects injects into the tensor product
of the two images.
\bigskip

(1)  $ F_{\gamma } X \bigotimes Y \subset F_{\gamma } X \bigotimes
F_{\gamma } Y$.

\bigskip

Equation 1 is expressed by mathematicians by saying that the functor is CoLax.

However, the injection is improper, in the sense that a dirac delta
function is an improper object of $L^2(R)$. This is because
REP(SL(2,C) contains representations labelled by a continuous
parameter, 
and the tensor product of two
representations is a direct integral in the sense of Mackey or Gelfand.  

This fact is physically important in the interpretation of the model
because it connects the discrete spectrum of areas in loop quantum
gravity or 3d TQFT with the continuum of representations in spacetime
models. The geometrical data on tetrahedra in the EPRL model
correspond to intertwiners in SO(3,R), lifted by the time functor, and
not to general SL(2,C) intertwiners.

Since the image category consists of infinite dimensional
representations, there is no hope of the trace in the domain category
going over via the functor to the range. However, there is a
renormalizable trace which is well defined on an amplee set of
diagrams in the domain category.

This renormalizable trace is just the multiple integral over SL(2,C) [3,4], 
which played a crucial role in the finiteness of the BC model, and
goes over to the new model. The renormalization just consists in
dropping one of the integrations (it doesnt matter which). The
finiteness of the resulting integral expression is the key to the
success of both models.

The ample set of diagrams on which the renormalized trace is finite
includes the free graph on 4 or 5 vertices, which represent the
tetrahedron and 4-simplex, and includes any diagram obtained by adding
edges to any diagram already in the ample set \cite{JJ}.

We can formalise these properties by defining a {\bf CLaren functor}
(Co-Lax, amply renormalizable) between
two tensor categories as one with an inclusion as in equation 1 for
any pair of objects X,Y in the first category, and a regularizable
trace for the images under the functor of an ample family of diagrams in the 
second category as defined above. 

Clearly, any CLaren functor can be used to construct a model analogous
to the EPRL model. It remains to be seen if other such models can be
physically useful, for example in unified theories.

\bigskip

{\bf 3D TQFT and the EPRL model.}

\bigskip

The formalization of the EPRL model we have proposed above connects it
with the general program of categorical construction of TQFT's. The
time 
functor connects two tensor categories, each of which
can be used to construct a state sum model in its own right, giving
CSW theory \cite{W} and BF theory \cite{CY}respectively. 

It has been suggested \cite{K} that the Chern-Simons Lagrangian could be the
state of the entire universe for quantum gravity. The parameter q in
the corresponding quantum group is associated with the cosmological
constant, which is not believed to be zero in the real world.

Now let us see if we can implement this idea by relating the 
EPRL model to the 2+1 dimensional
TQFT constructed from the quantum group $U_q(SO(3))$.

The Hilbert space associated to a triangulated 3-manifold in the EPR
paper is generated
by all spin networks one can associate with the tetravalent graph with
vertices on the tetrahedra of the manifold and edges passing through
its faces. 

It is a standard construction in Mathematics to associate a 4- holed
sphere to each tetrahedron of a 3-manifold, and glue them along their
corresponding boundaries at each face. Since this is the technique used to
associate a Heegaard splitting to a triangulation, let us call it the
{\bf Heegaard surface} of the triangulation. 

If we consider the
technique used to assign a vector space to a closed surface in the
construction of a 3d TQFT in \cite{C}, and apply it to the surface
we have just associated to a triangulation, we see that it corresponds
to the Hilbert space of the EPRL model, except that we are labelling
with representations of the quantum group rather than a classical one.

To see this, note that a trinion decomposition of the 4-holed sphere
just involves cutting it once, and that the representation on the cut
corresponds to the intertwiner in the EPRL picture.

Using  the CSW TQFT, we have, in addition to analogs of the EPRL Hilbert spaces of
triangulations, a consistent family of natural maps between them
corresponding to cobordisms between the surfaces. Formally, we can construct these 
by using the Chern-Simons path integral \cite{W}. For example, if we
change a triangulation by subdividing one tetrahedron into four, it is
easy to arrange matters so that one surface is entirely inside the
other. The region between the two surfaces would be a cobordism between
them, and hence give a linear map connecting them by the techniques of
3D TQFT in \cite{C}.

This gives us a map between the Hilbert spaces assigned to the Heegaard 
surfaces 
corresponding to the two
triangulations. 

Now we have the possibility of interpreting the Hilbert spaces on
different triangulations in a very different way. If there were no maps
relating them, we would have no choice but to sum them as in a GFT
picture, thus losing all the finiteness of the theory. 

Both the physical problems discussed in the introduction and the
mathematical foundations of the continuum appear in a different light
with this extra structure. The Hilbert spaces associated with
different triangulations of a region are connected by linear maps
which are consistent whenever cobordisms are composed, forming a
plausible starting point for an intrinsic description of a quantum
region. 

The linear map we have cited between the Hilbert space on the Heegaard surface 
of one triangulation and the one on a refinement of it would assign probabilities on 
labellings of any composite face to any labelling of the elementary faces of a triangulation.
This would allow us to study the case of larger regions in the
model on a given triangulation, which we suggested was the natural setting for
the classical limit in problem 1 above.

In the case of a bounded region, we shall now see that a quantum
version of holography emerges from our picture.

\bigskip

{\bf Holography in the new model}.

\bigskip

Now let us see if our new tool allows us to recover the Bekenstein
bound in the context of the new model.

It is not possible to do this rigorously, because at the moment we do
not have a theory to explain how light or matter propagates in the 
background of
a quantum gravitational model.  The Bekenstein bound \cite{Be} describes how
much information can pass from a bounded region to its environment.

However, there is a very reasonable hypothesis we can make. Suppose we
have a bounded region R in space whose boundary S consists of a finite
number of triangles, labelled with representations of SO(3,R). Let us
assume that each triangle is like an observer who sees a fuzzy sphere
with dimension corresponding to the representation on it.

In other words, let us take it that the information which can pass
through the quantum region corresponding to each triangle is described
by the underlying Hilbert space of the representation on it.

The total information which can pass through S would
then be described by the Hilbert space $H_S$ of the tensor product of all
the 
representations on the triangles
of S.

(Assuming that the quantum description of S is given by a single
(enormous) set of Planck scale triangles with fixed representations is
a simplifying assumption. In what follows we also assume that any
processes we study inside R have a negligible effect on S, so the
quantum geometry of S can be taken as fixed. We believe these
assumptions can be relaxed at the cost of greater technical
difficulty.)

If we identify the representations on the triangles with lines
crossing S in the loop gravity picture, which add units to the area 
\cite{Sm}, we find that the entropy of
the boundary of a region is proportional to the area of its boundary
(the dimension of $H_S$ grows exponentially with the number of representations, since it is a tensor product). This gives the
Bekenstein bound, up to a normalisation which we will not consider
here.

Information cannot flow out of R without crossing S. So the dimension
of $H_S$ as
defined above, is a bound for the information which the outside world
can access about the internal geometry of R.

Now imagine we have a triangulation T of R much less fine than the
Planck scale, so that each boundary face of T on S contains many
fundamental faces of S. This would be a realistic description of any
experiment to study the geometry of R, which would have much less than
Planck scale sensitivity. Assign to it the Heegaard surface of T,
$\Sigma _T$ . This
is a surface with one boundary circle for each boundary triangle of
T. Label each boundary triangle of T with the tensor product of all
the representations on fundamental triangles of S contained inside
it. (We assume the boundary of T is a coarsening of the triangulation
on S.)

Now the techniques of 3D TQFT allow us to assign to $\Sigma _T$ a
Hilbert space $H_T$. This is a quantum deformed version of the EPR
Hilbert space, restricted to a bounded region. We can think of the
q-deformation as accounting for the cosmological constant. States of
$H_T$ can be interpreted as superpositions of semiclassical geometries
of T using the methods of EPRL.

The region between $\Sigma _T$ and S gives us a cobordism between
them, so the methods of 3D TQFT give us a map 

\bigskip

$L_T : H_T \rightarrow H_S $.

\bigskip

Any external observer to R could only observe information about
the geometry of T which was contained in the image of $L_T$. Since
TQFT takes composites of cobordisms to composites of linear maps, the
filtration of observable information by the $L_T$'s would be compatible
with the maps linking the different triangulations described above.

We refer to the process of mapping the Hilbert spaces on the Heegaard
surfaces of the triangulations of R to $H_S$ as {\bf Localization at the boundary}, by
analogy with an important operation in model category theory \cite{Book}. 

In order to make predictions about observations an external observer
might make on the geometry of T, we only need operators defined on its
localisation, i.e. on $Im(L_T)$. A sufficiently dense triangulation
could exhaust $H_S$, so that calculations in a model based on it would
give the exact theory.

In the localisation picture, different triangulations can give
complementary pictures, rather than just corresponding to orthogonal
subspaces of a larger Hilbert space. 

The picture of apparent quantum geometry embedded in the CSW TQFT 
we have outlined here can be directly probed by doing well 
understood calculations \cite{C}.

\bigskip

{\bf Time evolution and localisation}.

\bigskip

In order for the localisation picture to correspond to a physical
theory, it must be consistent with the time evolution of the EPRL
model. 

We conjecture that this is so. There should be a way to correlate the
time evolution of a bounded spacetime region in the EPRL model on two different
four dimensional triangulations by using BF theory of the group
SL(2,C). The time functor should make this consistent with the
localisation we have described for space. This remains to be defined and 
demonstrated, although we have a natural tool because Heegaard decompositions 
are
special cases of handlebodies, which exist in all dimensions. 

We expect this to work out, because of the observation of Kodama \cite{K} 
that the Chern-Simons state is a solution to the equations of motion for quantun gravity.

Thus states with holography should evolve into states with holography.
\bigskip

{\bf On the appearance of quantum spacetime regions.} 

\bigskip

In \cite{Cr2}, we proposed that a spacetime region would appear as a
thickening of classical spacetime, because regions in different
superimposed metrics would not appear to be in the same place as an
observer changed the angle from which it observed them.

It is natural to implement this in the EPRL model by regarding the
triangulations with different labelling representations as different
places. Without the localisation maps, this would result in many
parallel copies of the spatial region. Localization allows us to
make linear superpositions of the images of the differently labelled
triangulations in $H_S$, thus filling in between them.

The picture of a ``total space'' of an EPRL spin foam which results
from this idea is very similar to constructions which are well known
in model category theory. For example, in the Sullivan model \cite{S}, a
simplicial set is described by the graded algebra of polynomial
differential forms on it. This can be converted back to a simplicial
complex by taking simplices with all possible forms on them. The
result is a thickened up version of the original space, which has the
same topology. This is a standard construction of a fibrant model for
a space.

The total space of a spin foam model can be thought of as a
quantization of the Sullivan model, in that the representations on the
faces in BC or EPRL are quantizations of the differential forms which
describe their classical geometry. 

Fibrant models are the key technical tool in model category
theory. They allow us to replace point sets with complexes in
algebraic and differential topology. The idea that quantum
descriptions of space and spacetime have the structure of fibrant
models is deeply suggestive that quantum gravity may lead to a
pointless foundation of geometry. Perhaps if we could see down near
the Planck scale, spacetime would actually look like a fibrant model,
with many parallel worlds with different shapes which blended into one
another because of the Planck scale limitations on information flow.

\bigskip

{\bf Unification}

\bigskip

The construction of quantum gravity from CSW theory by means of a
CLaren functor suggests that one could search for unified models
including chiral fermions, Yang Mills fields etc. by searching among
the 3D TQFTs. Since construction of TQFT in 3D out of tensor
categories is well understood, this might be relatively easy.

\end{document}